      \documentclass[prb,twocolumn,showpacs,
      preprintnumbers,amsmath,amssymb]{revtex4}

\usepackage{graphicx}
\usepackage{dcolumn}
\usepackage{bm}

\begin{document}


\title{Time-reversal symmetry breaking surface states of $d$-wave superconductors 
induced by an additional order parameter with negative $T_c$
}

\author{Takeshi Tomizawa,$^{1,2}$}
\author{Kazuhiro Kuboki$^1$}%
\email{kuboki@kobe-u.ac.jp}
\affiliation{%
$^1$Department of Physics, Kobe University, \\
Kobe 657-8501, Japan
}%
\affiliation{%
$^2$Department of Physics, Nagoya University, \\
Nagoya 464-8602, Japan
}%

\date{\today}

\begin{abstract}
Surface states of $d_{x^2-y^2}$-wave superconductors are studied using the  
Ginzburg-Landau (GL) theory. For a [110] surface it has been known that 
the time-reversal symmetry (${\cal T}$) breaking surface state, 
($d \pm is$)-wave state, can occur if the bare transition temperature of 
the $s$-wave order parameter (OP) is positive. 
We show that even if this bare $T_c$ is negative, it is possible to break 
${\cal T}$ because the coupling to the spontaneously generated magnetic 
field may induce the $s$-wave OP. The ${\cal T}$-breaking state is favored  
when the GL parameter $\kappa$ is small.
\end{abstract}

\maketitle

\section{Introduction}

Superconducting (SC) states of high-$T_c$ cuprates are known to have 
$d_{x^2-y^2}$-wave symmetry. \cite{Harlingen,Tsuei}
Since the pair wave function of such an unconventional SC state has 
strong angular dependence, the effects of the presence of surfaces, impurities 
are different from those in conventional $s$-wave superconductors. 
For example, it is possible to break the time-reversal symmetry (${\cal T}$) 
near a surface or a Josephson junction by inducing the second component of the 
SC order parameter 
(OP)\cite{Tsuei,Sig,TK,Vol,SRU,SBL,KS1,KS2,MS1,MS2,Forgel,Covin,HWS}
with a nontrivial phase difference between the two OPs.  
In the case of a Josephson junction it may occur when the surface has [110] 
orientation, because the second SCOP induced by the tunneling process 
can have phase difference $\pm \pi/2$ leading to a ${\cal T}$-breaking 
state.\cite{KS1,KS2} 
For a [110]  surface faced to a vacuum the necessary condition to 
break ${\cal T}$ seems to be that the bare transition temperature ($T_c$) of 
the second OP is positive. \cite{SBL,MS1,MS2,Forgel,HWS}

In this paper we examine the possibility to have a ${\cal T}$-breaking surface 
state near the [110] surface of a $d_{x^2-y^2}$-wave superconductor 
when the bare $T_c$ of the additional OP is negative, namely, 
the second OP will not occur in the bulk even at zero temperature.   
We take an $s$-wave SCOP as the second component,  
since $d_{x^2-y^2}$-wave and extended $s$-wave symmetries
are natural candidates for superconducting states in the 
models with nearest-neighbor interactions ({\it e.g.}, the $t-J$ model). 
We will show that this kind of ${\cal T}$-violation is possible, 
and that both the SCOPs and the magnetic field (vector potential) 
should be treated self-consistently in order to describe this situation correctly. 
It also turns out that the ${\cal T}$ violation may occur for a 
relatively small GL parameter $\kappa$ ({\it i.e.}, of the order of 10),    
when $T_c$ of the second OP is negative. 
Then the present mechanism may not be relevant to the 
${\cal T}$-violation in hole-doped cuprates in which $\kappa \sim 100$. 
However, we expect the surface states of electron-doped cuprates may be 
described by the present theory, because some of the latter systems  
have much smaller $\kappa$ values.\cite{Edope1,Edope2} 

\section{Ginzburg-Landau equation}

 We consider a superconductor with tetragonal symmetry and assume only a 
$d_{x^2-y^2}$-wave SCOP, $\Delta_d$, is present in the bulk. 
An $s$-wave SCOP, $\Delta_s$, is taken into account as a possible second 
component when $\Delta_d$ is suppressed near the surface. 
For such a system the Ginzburg-Landau (GL) free energy is given as\cite{Sig} 
\begin{equation}\begin{array}{rl}
{\cal F} = & \displaystyle \int d{\bf r} \Big[\sum_{\mu=d,s} 
\Big(\alpha_\mu |\Delta_\mu|^2 
+ \frac{\beta_\mu}{2} |\Delta_\mu|^4 + K_\mu |{\bf D}\Delta_\mu|^2\Big)  \\
+ & \displaystyle  \gamma_1 |\Delta_d|^2|\Delta_s|^2 
+ \gamma_2 \{\Delta_d^2(\Delta_s^*)^2 + (\Delta_d^*)^2\Delta_s^2\}  \\
+ & \displaystyle K_{ds} \Big\{(D_x\Delta_d)(D_x\Delta_s)^* 
- (D_y\Delta_d)(D_y\Delta_s)^* + c.c \Big\} \\
+  & 
\displaystyle 
\frac{1}{8\pi} (\nabla \times {\bf A})^2 \Big]
\end{array}\end{equation} 
where ${\bf A}$ is the vector potential and 
$\displaystyle {\bf D} = {\nabla} - (2\pi i/\Phi_0){\bf A}$ 
is the gauge invariant gradient with 
$\Phi_0=hc/2e$ being the magnetic flux quantum. 
Coefficients $\alpha_\mu (\propto T-T_{c\mu})$, $\beta_\mu$, $K_\mu$, 
$\gamma_1$, $\gamma_2$ and $K_{ds}$ are real, and we assume $T_{cd} >0$, 
while $T_{cs}$ can be both positive and negative. 
The $\gamma_2$ is one of the terms which 
determine the relative phase of OPs, 
$\phi_{ds} (\equiv \phi_d-\phi_s; \Delta_\mu=|\Delta_\mu|\exp(i\phi_\mu)$).   
We take $\gamma_2 > 0$, because this choice would lead to 
the $(d \pm is)$-state ($\phi_{ds} = \pm\pi/2$) instead of the $(d \pm s)$-state 
($\phi_{ds} = 0, \pi$).  In the former case the nodes of the $d$-wave state are 
removed and the more condensation energy can be gained. 
It is also to be noted that $\gamma_1 \pm 2\gamma_2$ is positive in usual 
weak-coupling model, since two OPs compete each other. 
Now we rewrite ${\cal F}$ in the dimensionless 
unit\cite{Fetter} to see the parameter dependence of the model more clearly, 
\begin{equation}\begin{array}{rl}
{\cal F} = & \displaystyle \frac{H_c^2\xi_d^3}{4\pi}\int d{\bf r} 
\Big[-|\eta_d|^2 + \frac{1}{2} |\eta_d |^4 \
+ |{\bf {\tilde D}}\eta_d|^2 \\
& \displaystyle 
+ {\tilde \alpha}_s|\eta_s|^2 + \frac{{\tilde \beta}_s}{2} |\eta_s |^4 
+ {\tilde K}_s|{\bf {\tilde D}}\eta_s|^2 \\
& \displaystyle + {\tilde \gamma}_1|\eta_d|^2|\eta_s|^2 
+ {\tilde \gamma}_2 \Big(\eta_d^2(\eta_s^*)^2 
+ (\eta_d^*)^2\eta_s^2\Big) \\ 
& \displaystyle 
+ {\tilde K}_{ds} \Big(({\tilde D}_x\eta_d)({\tilde D}_x\eta_s)^* 
- ({\tilde D}_y\eta_d)({\tilde  D}_y\eta_s)^* + c.c.\Big) \\ 
& \displaystyle 
+ (\nabla \times {\bf a})^2\Big], 
\end{array}\end{equation} 
where $\eta_\mu = \Delta_\mu/\Delta_0$ ($\mu=d,s$) with 
$\Delta_0 = \sqrt{|\alpha_d|/\beta_d}$ being the bulk $d$-wave OP.  
${\bf r}$ was rescaled using the coherence length for the $d$-wave OP, 
$\xi_d$ ($=\sqrt{K_d/|\alpha_d|}$), as ${\bf r} \to {\bf r}/\xi_d$, and 
${\bf {\tilde D}} \equiv \nabla - i{\bf a}/\kappa$.   
Here ${\bf a} = {\bf A}/(\sqrt{2}H_c\xi_d)$, and the magnetic field  
is measured in units of $\sqrt{2}H_c$, 
where $H_c = \sqrt{4\pi\alpha_d^2/\beta_d}$ is the thermodynamic critical field. 
$\kappa=\lambda_d/\xi_d$  is the GL parameter 
with $\lambda_d = \phi_0/(2\sqrt{2}\pi H_c\xi_d)$ being the penetration 
depth for the bulk $d$-wave superconductor. 
The parameters in Eq.(2) are defined as ${\tilde \alpha}_s=\alpha_s/|\alpha_d|$, 
${\tilde \beta}_s=\beta_s/\beta_d$, ${\tilde K}_s=K_s/K_d$, 
${\tilde \gamma}_1=\gamma_1/\beta_d$, 
${\tilde \gamma}_2=\gamma_2/\beta_d$ and 
${\tilde K}_{ds}=K_{ds}/K_d$. 

Usually the surface effect is described by the second-order surface GL 
free energy,  
$
\displaystyle 
{\cal F}_{\rm sf} =  \int_{\rm sf} dS \sum_{\mu,\nu=d,s} 
g_{\mu\nu} \eta_\mu^*\eta_\nu,  
$
where integration is carried out on the surface. 
\begin{figure}[htb]
\begin{center}
\includegraphics[width=4.0cm,clip]{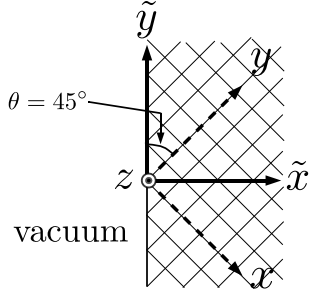}
\caption{Schematic of a [110] surface of a $d_{x^2-y^2}$-wave 
                superconductor with tetragonal symmetry. $x$ and $y$ are 
				parallel to the crystal $a$ and $b$ axes, respectively.}
\end{center}
\end{figure}
Using the symmetry argument we find $g_{ds} = g_{sd} = g_0\cos2\theta$ 
where $\theta$ is the angle between the surface 
and the crystal $a$-axis with $g_0$ being a constant. 
This term could also determine $\phi_{ds}$, and it leads to the $(d\pm s)$-state 
in the case of a [100] surface ($\theta=0$), since the $\gamma_2$ term 
is higher order than the $g_{ds}$ term. However, $g_{ds}$  
vanishes for a [110] surface ($\theta=45^\circ$) which we consider 
in the following. The $g_{\mu\mu}$ term will represent the suppression 
of $\eta_\mu$ near the surface. 
Instead of using $g_{dd}$ we impose the condition $\eta_d = 0$ at 
the [110] surface,  because the $d_{x^2-y^2}$-wave SCOP should vanish there. 
Since the $s$-wave SCOP is only little affected by the presence of the surface, 
we take $g_{ss}=0$. 
(In numerical calculations we have checked that taking small positive 
$g_{ss}$ will not change the results qualitatively.) 
In order to consider the [110] surface we transform the coordinate system,   
$(x,y,z) \to ({\tilde x}, {\tilde y},z)$. 
Here $x$ ($y$) is parallel to the crystal $a$ ($b$) axis  
($z$ is parallel to the surface), and ${\tilde x}$ and ${\tilde y}$ axes are 
perpendicular and parallel to the surface, respectively. (See Fig.1.) 
In the free energy density only the ${\tilde K}_{ds}$ term is changed under 
this transformation to  
\begin{equation}
\displaystyle 
\frac{2{\tilde K}_{ds}}{\kappa} a_{\tilde y} 
{\rm Im} (\eta_s^*\partial_{\tilde x} \eta_d 
- \eta_d \partial_{\tilde x} \eta_s^*) 
\end{equation}
where we have assumed that the system is uniform along the surface,  and 
the gauge freedom was taken as 
${\bf a} = a_{\tilde y}({\tilde x}){\bf e}_{\tilde y}$. 

The expression for the supercurrent is obtained by varying the electronic 
part of ${\cal F}$ ({\it i.e.}, except the last term) with respect to ${\bf a}$. 
Since the surface is faced to the vacuum, 
the ${\tilde x}$ component, $J_{\tilde x}$, should obviously vanish.
 (We have numerically checked that $J_{\tilde x}$ actually vanishes.) 
The ${\tilde y}$ component, $J_{\tilde y}$, and that in the the dimensionless unit,   
$j_{\tilde y}$, are given as 
\begin{equation}\begin{array}{rl}
j_{\tilde y} = 
& \displaystyle J_{\tilde y}\big/\big(\frac{\sqrt{2}H_cc}{\xi_d}\big) \\
= & -  \displaystyle \frac{1}{4\pi} 
\Big[\frac{1}{\kappa^2} a_{\tilde y} 
(|\eta_d|^2 + {\tilde K}_s |\eta_s|^2) 
\\ 
& \displaystyle 
+  \frac{{\tilde K}_{ds}}{\kappa} 
{\rm Im} (\eta_s^* \partial_{\tilde x} \eta_d 
- \eta_d \partial_{\tilde x} \eta_s^*)\Big]. 
\end{array}
\end{equation}

\section{Surface state and spontaneous current}

We numerically solve the problem by employing the quasi-Newton 
method\cite{NR} to minimize the free energy ${\cal F}$ 
under the condition $\eta_d({\tilde x}=0)=0$. 
We minimize ${\cal F}$ with respect to all variables, {\it i.e.},  
$\eta_d$, $\eta_s$ and $a_{\tilde y}$.  
Note that the Maxwell's equation is taken into account in this procedure, 
and we call this as "fully self-consistent calculation". 
For the sake of comparison we will also show the results by treating only 
$\eta_d$ and  $\eta_s$ self-consistently. 

First let us consider the case of ${\tilde \alpha}_s  <  0$ ({\it i.e.}, $T < T_{cs}$). 
In this case, we would get finite $\eta_s$ if $\eta_d$ were absent. 
However, for $T_{cd} > T_{cs}$ the stability condition for $\eta_s$ in the bulk is 
given as,  
$
{\tilde \alpha}_s + ({\tilde \gamma}_1-2{\tilde \gamma}_2)|\eta_d|^2 < 0,
$
so the transition temperature of $\eta_s$ is lower than the bare one, $T_{cs}$,  
and $\eta_s$ would be totally suppressed if $T_{cd} \gg T_{cs}$. 
Near the surface or impurities the situation can be different.
There $\eta_s$ may be finite because the dominant SCOP, $\eta_d$, is suppressed.  
In Fig.2 the spatial variations of the SCOPs near the surface are shown.
$\eta_s$ gets finite near the surface while $\eta_d$ is suppressed. 
The relative phase $\phi_{ds}$ will be 
determined by ${\tilde \gamma}_2$ and ${\tilde K}_{ds}$ terms, and 
the former favors $\phi_{ds} = \pm \pi/2$ as mentioned. 
From Eq.(3) we see that the ${\tilde K}_{ds}$ term also favors 
$\phi_{ds} = \pm \pi/2$, and  $a_{\tilde y}$ will be spontaneously generated.  
(We take $\eta_d$ to be real and $a_{\tilde y}=0$ in the bulk, 
{\it i.e.}, ${\tilde x} \to \infty$.) 
Numerical calculations show that $\eta_d$ is real for all ${\tilde x}$, 
and that $\phi_{ds} = \pm \pi/2$ where $\eta_s$ is finite. 
This indicates that a ${\cal T}$-violating ($d+is$)-wave surface state 
with a spontaneous magnetic field $b_z$ ($= \partial_{\tilde x} a_{\tilde y}$) 
and a supercurrent $j_{\tilde y}$ occurs near the surface. 
The spatial distributions of $b_z$  and $j_{\tilde y}$ are presented in Fig.3. 

In order to see the role played by the vector potential, 
we investigate the same problem by setting $a_{\tilde y}=0$ everywhere.
Namely we treat only SCOPs self-consistently. 
When $a_{\tilde y}$ is set to zero, the spontaneous current 
$j_{\tilde y}$ has contributions from only the spatial variations of SCOPs 
({\i.e.}, the last line of Eq.(4)), 
and we calculate the magnetic field 
from $j_{\tilde y}$ using Maxwell's equation, 
$j_{\tilde y}({\tilde x}) = 
- \frac{1}{4\pi} \frac{\partial b_z({\tilde x})}{\partial {\tilde x}}$. 
For ${\tilde \alpha}_s < 0$, the results for the SCOPs look similar 
as in the fully self-consistent calculations. 
The ${\cal T}$-breaking $(d + is)$-state occurs as shown in Fig.2. 
On the contrary, the behaviors of $b_z$ and $j_{\tilde y}$ are different  
in that $j_{\tilde y}$ always has the same sign, and that $b_z$ is a 
monotonous function of ${\tilde x}$. 
These results are not correct even qualitatively as well as in a quantitative sense. 
Integration of the Maxwell's equation with the boundary condition 
$b_z(\pm \infty) = 0$ leads to 
$
\int_{-\infty}^\infty d{\tilde x} j_{\tilde y}({\tilde x}) = 0,  
$
implying that the averaged current should vanish. \cite{KS2}
This is the case for the fully self-consistent calculation 
but not in the case where the magnetic field is not treated self-consistently, 
because of the absence of the screening effect in the latter.

\begin{figure}[htb]
\begin{center}
\includegraphics[width=6.0cm,clip]{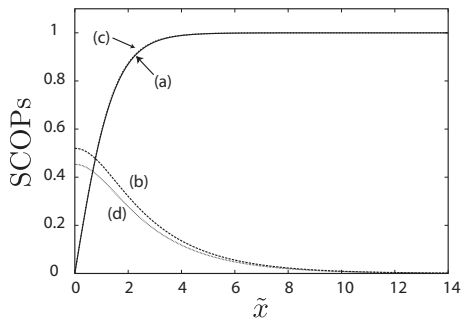}
\caption{Spatial variations of SCOPs for ${\tilde \alpha}_s=-0.2, 
     {\tilde \beta}_s=0.2, {\tilde K}_s=0.5$, ${\tilde \gamma}_1=0.5, 
	 {\tilde \gamma}_2=0.1, {\tilde K}_{ds}=0.3$ and $\kappa=16$. 
     Note that all SCOPs are normalized by the bulk $d$-wave OP, 
	 and ${\tilde x}=0$ corresponds to the surface faced to the vacuum.  
	 (a) Re$\eta_d$ and (b) Im$\eta_s$ in the fully self-consistent calculation.
	 (c) Re$\eta_d$ and (d) Im$\eta_s$ in the simplified one without treating 
	 $a_{\tilde y}$ self-consistently. 
	 }
\end{center}
\end{figure}
\begin{figure}[htb]
\begin{center}
\includegraphics[width=7.5cm,clip]{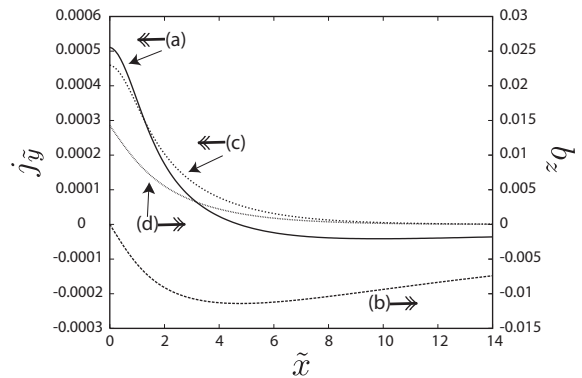}
\caption{Spatial variations of $b_z$ and $j_{\tilde y}$. Parameters used are  
     the same as in Fig.2. 
	 (a) $j_{\tilde y}$ and (b) $b_z$ in the fully self-consistent calculation.
	 (c) $j_{\tilde y}$ and (d) $b_z$ in the simplified one without treating 
	 $a_{\tilde y}$ self-consistently. Note $b_z$ and $j_y$ are in the 
	 dimensionless unit.}
\end{center}
\end{figure}

Next we consider the case of ${\tilde \alpha}_s >0$,  {\it i.e.}, $T > T_{cs}$. 
Note that $T_{cs}$ may be negative,  in which case $\eta_s$ will not 
occur in the bulk at $T=0$ even when $\eta_d$ is absent. 
The results for the SCOPs are depicted in Fig.4. 
(Here the GL parameter is taken to be $\kappa=16$.) 
It is seen that finite Im($\eta_s$) is obtained, though we naively 
expect $\eta_s =0$. 
This is because the ${\tilde K}_{ds}$ term couples 
$\partial_{\tilde x}$Re($\eta_d$) bilinearly to $a_{\tilde y}$Im($\eta_s$).
It may induce the state with Im$(\eta_s)\not=0$ and $b_z\not=0$, but 
the state with $\eta_s=0$ and $b_z=0$ may also be a self-consistent solution. 
Numerical calculations show that the former one has the lower energy,
and thus the time-reversal symmetry is violated spontaneously.  
Here $|{\tilde \alpha}_s|$, ${\tilde \beta}_s$ and ${\tilde K}_s$ 
were taken to be much smaller than those in Fig.2. 
Otherwise the ${\cal T}$-violation will not occur, 
because these terms cost the energy for ${\tilde \alpha}_s > 0$ and 
the energy gain is solely coming from the ${\tilde K}_{ds}$ term. 
The spatial variations of $b_z$ and $j_{\tilde y}$ are shown in Fig.5. 

\begin{figure}[htb]
\begin{center}
\includegraphics[width=6.0cm,clip]{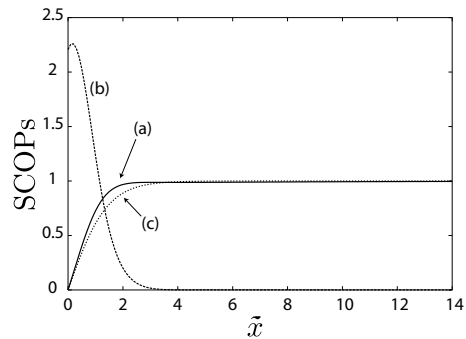}
\caption{Spatial variations of SCOPs for ${\tilde \alpha}_s=0.01, 
     {\tilde \beta}_s=0.01, {\tilde K}_s=0.04$, ${\tilde \gamma}_1=0.5, 
	 {\tilde \gamma}_2=0.1, {\tilde K}_{ds}=1.0$ and $\kappa=16$. 
	 (a) Re$\eta_d$ and (b) Im$\eta_s$ in the fully self-consistent calculation.
	 (c) Re$\eta_d$ in the simplified one without treating 
	 $a_{\tilde y}$ self-consistently. 
	 }
\end{center}
\end{figure}
\begin{figure}[htb]
\begin{center}
\includegraphics[width=7.5cm,clip]{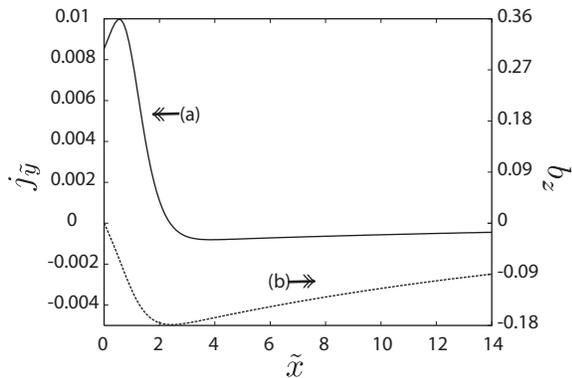}
\caption{Spatial variations of $b_z$ and $j_{\tilde y}$. Parameters used are  
     the same as in Fig.4.  (a) $j_{\tilde y}$ and (b) $b_z$ in the fully 
	 self-consistent calculation. 
	 }
\end{center}
\end{figure}

In the case of ${\tilde \alpha}_s > 0$, 
the results with or without treating the vector potential self-consistently are 
completely different. 
If we do not take into account the $a_{\tilde y}$ term, $\eta_s$ 
will never appear, since there is no mechanism to derive finite $\eta_s$. 
Thus neither the spontaneous current nor the spontaneous field can occur.  
It implies that the ${\cal T}$-violation near the surface cannot be described 
in this kind of simplified treatment for the superconductors in which 
the second SCOP has negative $T_c$. 

In order to see the dependence on $\kappa$ we show 
the results for a larger $\kappa$ ($\kappa=19$) in Fig.6 and 7. 
It is seen that $|\eta_s|$, $|b_z|$ and $|j_{\tilde y}|$ are much smaller than 
those for $\kappa=16$.
This $\kappa$ dependence can be understood as follows. 
$\eta_d$ is suppressed in the region near the surface (${\tilde x} \lesssim \xi_d$), 
and $\eta_s$ and $a_{\tilde y}$ would be finite there if ${\cal T}$ is broken. 
On the other hand the magnetic field $b_z$ would 
be finite in the region ${\tilde x} \lesssim \lambda_d$. 
When $\kappa$ is large, the loss of energy due to finite $b_z$ 
in the large region ($\xi_d  \lesssim {\tilde x} \lesssim \lambda_d$) 
overwhelms the energy gain coming from the ${\tilde K}_{ds}$ term which acts 
only in the small region ${\tilde x} \lesssim \xi_d$. 
Thus for large $\kappa$ the ${\cal T}$-violation is not favored. 
If the larger value of ${\tilde K}_{ds}$ is taken, 
the ${\cal T}$-breaking state can occur for larger $\kappa$.
But the natural assumption seems to be $K_{ds} \leq K_d$ (${\tilde K}_{ds} \leq 1$), 
so that the ${\cal T}$-violation may occur for $\kappa$ of the order of 10. 
(On the contrary the ${\cal T}$-violation may occur for much larger $\kappa$
in the case of ${\tilde \alpha}_s < 0$, because the energy can be gained by not only 
${\tilde K}_{ds}$ but also ${\tilde \alpha}_s$ term.) 
It implies that the present mechanism may not be relevant to hole-doped 
cuprates in which $\kappa \sim 100$, but 
it may describe the surface states of 
electron-doped cuprates which have smaller $\kappa$.

\begin{figure}[htb]
\begin{center}
\includegraphics[width=6.0cm,clip]{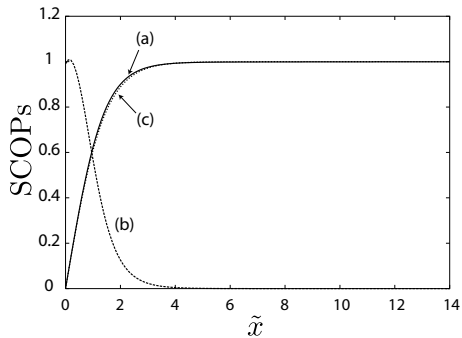}
\caption{Spatial variations of SCOPs. 
     Parameters used are the same as in Fig.4 except $\kappa=19$. 
	 (a) Re$\eta_d$ and (b) Im$\eta_s$ in the fully self-consistent calculation.
	 (c) Re$\eta_d$ in the simplified one without treating 
	 $a_{\tilde y}$ self-consistently. 
	 }
\end{center}
\end{figure}
\begin{figure}[htb]
\begin{center}
\includegraphics[width=7.5cm,clip]{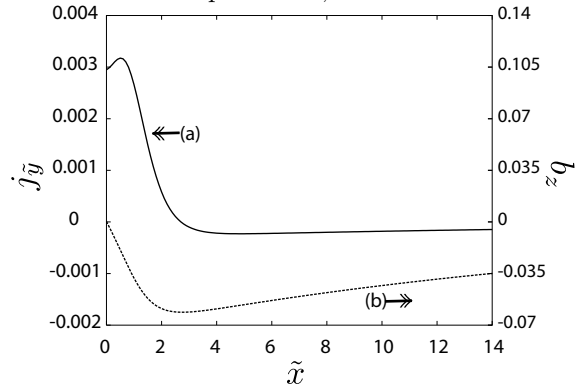}
\caption{Spatial variations of $b_z$ and $j_{\tilde y}$. Parameters used are  
     the same as in Fig.6. (a) $j_{\tilde y}$ and (b) $b_z$ in the fully 
	 self-consistent calculation. 
	 }
\end{center}
\end{figure}

If we assume $H_c=$1T, the maximum values of $|B_z|$ and 
$|J_{\tilde y}|$ are $2.5\times10^{-1}$T and $3.7\times10$A/cm$^2$, 
respectively, for $\kappa=16$. 
For $\kappa=19$ they are $8.6\times10^{-2}$T and $1.2\times 10$A/cm$^2$, respectively. 
These values rapidly decrease as $\kappa$ increases, 
and the ${\cal T}$-breaking state disappears as $\kappa$ exceeds 19  
for the parameters used here. If we compare these values
with experiments, it should be noted that surface roughness will reduce 
$|B_z|$ and $|J_{\tilde y}|$, because ${\cal T}$-violation is most favored 
in the case of $\theta=45^\circ$.\cite{Forgel} 
(When $\theta \not= 45^\circ$, $g_{ds}$ will be finite and 
the ${\cal T}$-violation is not favored.)

\section{summary}

We have examined the role played by the vector potential 
concerning the occurrence of surface states with 
spontaneously broken ${\cal T}$ in $d_{x^2-y^2}$-wave superconductors. 
It has been known that the ${\cal T}$-breaking state may naturally 
appear if the bare $T_c$ of the additional OP is positive. 
For the Josephson junction composed of 
$d_{x^2-y^2}$-wave and other superconductors, tunneling may induce 
second component of SCOP and thus ${\cal T}$ may be broken.
In these cases the ${\cal T}$-breaking states may be described 
without treating the vector potential self-consistently. 
In this paper it was shown that the surface state of a 
$d_{x^2-y^2}$-wave superconductor may break ${\cal T}$ even when 
the bare $T_c$ of the second SCOP is negative. 
However, to describe this situation correctly not only the SCOPs but also 
the vector potential must be treated on an equal footing. 
In the present mechanism the ${\cal T}$-violation may occur for rather small 
values of the GL parameter $\kappa$ ($\lesssim 20$), so that it may not 
be relevant to hole-doped cuprates. 
We expect the present theory may be used to describe the surface states 
of electron-doped high-$T_c$ cuprates, because their $\kappa$ are much smaller 
than those of hole--doped systems.

%



\begin{references}

\bibitem{Harlingen} D. J. van Harlingen, Rev. Mod. Phys. {\bf 67}, 515 (1995). 

\bibitem{Tsuei} C. C. Tsuei and J. R. Kirtley, Rev. Mod. Phys. 
{\bf 72}, 969 (2000) .

\bibitem{Sig}  M. Sigrist, Prog. Theor. Phys. {\bf 99}, 899 (1998).

\bibitem{TK} S. Kashiwaya and Y. Tanaka, Rep. Prog. Phys. {\bf 63}, 1641 (2000).

\bibitem{Vol}  G.E. Volovik and L.P. Gor'kov, Zh. Eksp. Teor. Fiz. 
{\bf 88}, 1412  (1985)  [JETP,  {\bf 61}, 843  (1985) ].

\bibitem{SRU} M. Sigrist, T. M. Rice and K. Ueda,  Phys. Rev. Lett. 
{\bf 63}, 1727 (1989) . 

\bibitem{SBL}  M. Sigrist, D. B. Bailey and R. B. Laughlin,  Phys. Rev. Lett.  
{\bf 74}, 3249 (1995) .    

\bibitem{KS1}  K. Kuboki and M. Sigrist, J. Phys. Soc. Jpn.  {\bf 65}, 361  (1996). 

\bibitem{KS2}  K. Kuboki and M. Sigrist, J. Phys. Soc. Jpn.  {\bf 67}, 2873 (1998) .

\bibitem{MS1} M. Matsumoto and H. Shiba, J. Phys. Soc. Jpn. {\bf 64}, 3384 (1995) .

\bibitem{MS2} M. Matsumoto and H. Shiba, J. Phys. Soc. Jpn. {\bf 64}, 4867 (1995) .

\bibitem{Forgel} M. Fogelstr\"om, D. Rainer and J.A. Sauls, Phys. Rev Lett. 
{\bf 79}, 281 (1997).

\bibitem{Covin} M. Covington, M. Aprili, E. Paraoanu, L. H. Greene, F. Xu, 
J. Zhu and C. A. Mirkin, Phys. Rev Lett. {\bf 79}, 277 (1997) .

 \bibitem{HWS} C. Honerkamp, K. Wakabayashi and M. Sigrist, 
Europhys. Lett. {\bf 50}, 368  (2000). 

\bibitem{Edope1} L. F\'abrega, B. Mart\'inez, J. Fontcuberta, X. Obradors 
and S. Pi\~nol,  Phys. Rev. B {\bf 46}, 5581 (1992). 

\bibitem{Edope2}  F. Gollnik and M. Naito, Phys. Rev. B {\bf 58}, 11734 (1998). 

\bibitem{Fetter} See, for example, 
A. L. Fetter and P. C. Hohenberg, in {\it Superconductivity}, 
edited by R. D. Parks (Marcel Dekker, New York, 1969).  

\bibitem{NR} W. H. Press, S. A. Teukolsky, W. T. Vetterling and B. P. Flannery, 
{\it Numerical Recipes in Fortran }77, 2nd ed.  
(Cambridge University Press, Cambridge, 2003). 

\end{references}
\end{document}